# THE IONIZATION FRACTION IN DENSE CLOUDS

C. de Boisanger[1,2], F.P. Helmich[2], and E.F. van Dishoeck[2]

[1] *Commissariat à l'Energie Atomique, Centre d'Etudes de Bruyères-le-Châtel, Service PTN, F-91680 Bruyères-le-Châtel, France.*
[2] *Leiden Observatory, P.O.-Box 9513, 2300 RA Leiden, The Netherlands*

# Abstract

We present submillimeter observations of various molecular ions toward two dense clouds, NGC 2264 IRS1 and W 3 IRS5, in order to investigate their ionization fraction. Analysis of the line intensity ratios by the way of statistical equilibrium calculations allows determination of the physical parameters: $n(\mathrm{H_2}) \sim (1-2)\ 10^6\ \mathrm{cm^{-3}}$ and $T_{\mathrm{kin}} \sim 50-100$ K. Column densities and abundances are also derived. Together, the abundances of the observed ions provide a lower limit to the ionization fraction, which is $(2-3)\ 10^{-9}$ in both clouds. In order to better constrain the electron abundance, a simple chemical model is built which calculates the steady state abundances of the major positive ions, using the observed abundances wherever available. With reasonable assumptions, good agreement within a factor of two with the observations can be achieved. The calculated electron fraction is $x_e = (1.0 - 3.3)\ 10^{-8}$ in the case of NGC 2264 and $x_e = (0.5 - 1.1)\ 10^{-8}$ for W 3 IRS5. In the first case, the high abundance of $\mathrm{N_2H^+}$ requires a rather high cosmic ray ionization rate $> 10^{-16}\ \mathrm{s^{-1}}$, even if all nitrogen is assumed to be in gas phase $\mathrm{N_2}$. For W 3 IRS5, ionized metals such as $\mathrm{Fe^+}$ and $\mathrm{Mg^+}$ could provide 60% of the electrons.

# 1 Introduction

The ionization fraction of dense clouds is a basic parameter of interstellar physics, since it determines the efficiency with which magnetic fields couple with the gas and ultimately controls the rate of collapse and star formation. Moreover, electrons and ions play a major role in interstellar chemistry, and their abundances provide information on the nature and importance of sources which ionize the gas. It is well known that in the diffuse gas, interstellar photons with $\lambda > 912$ Å provide most of the ionization so that the electron fraction is of order $10^{-4}$, equal to the abundance of the major supplier $\mathrm{C^+}$. This fraction is thought to drop by several orders of magnitude inside dense clouds, where the ionization mostly results from cosmic ray interactions. The major ions here are thought to be $\mathrm{H_3^+}$, $\mathrm{H^+}$, $\mathrm{He^+}$, $\mathrm{HCO^+}$, $\mathrm{H_3O^+}$ and possibly metal ions (e.g. $\mathrm{Fe^+}$, $\mathrm{Mg^+}$). Unfortunately, except for $\mathrm{HCO^+}$, all of these species are very difficult to observe directly for spectroscopic reasons.

The first attempts to derive the electron abundance in dense clouds date back more than 15 years (Watson 1977; Guélin et al. 1977, 1982; Wootten et al. 1982). The basic idea was to use the observed $\mathrm{DCO^+}/\mathrm{HCO^+}$ ratio as a measure of the $\mathrm{H_2D^+}/\mathrm{H_3^+}$ ratio, which, together with a model for the deuterium fractionation and the $H_3^+$ recombination, gives a limit on the electron abundance. Typical electron fractions derived in this way were $10^{-8} - 5\ 10^{-7}$, with an order of magnitude variation from cloud to cloud (Langer 1985). However, this method was abandoned in 1984, when measurements by Adams & Smith (1984) found the dissociative recombination rate of $H_3^+$ with $e$ to be negligibly small, thereby invalidating the earlier determinations and raising the upper limits to $10^{-5}$. The $H_3^+$ rate has subsequently been a heated topic of discussion in astrochemistry for more than a decade, preventing further studies of the ionization fraction.

There are several reasons why a new attack on this problem is now both timely and warranted. First, the latest set of laboratory measurements on the $H_3^+$ dissociative recombination rate seem to converge to the same (relative rapid) value under interstellar conditions (Amano 1990; Canosa et al. 1992; Larsson et al. 1993), although some lingering uncertainties remain (Smith & Španel 1993). Second, many more ions can now be observed with the new sensitive, high frequency receivers: apart from $HCO^+$ and its isotopes, these include $HCS^+$, $N_2H^+$ and $H_3O^+$ (Phillips et al. 1992), whereas good upper limits on $H_2D^+$ exist for a number of regions (van Dishoeck et al. 1992). Third, the dissociative recombination rates of $HCO^+$, $HCS^+$, and $N_2H^+$ have also recently been measured in the laboratory (Rowe et al. 1993). Fourth, the excitation and optical depth of the lines in dense clouds is now much better constrained by observations of several submillimeter lines, thereby minimizing the uncertainties in abundance determinations and physical conditions.

We present here observations of various ions ($HCO^+$, $H^{13}CO^+$, $DCO^+$, $HCS^+$ and $N_2H^+$) in two different dense clouds, NGC 2264 and W 3 IRS5, in order to investigate their ionization fraction. The clouds are chosen to have a range in density and temperature (as determined from our earlier $H_2CO$ data) and to have complementary $H_3O^+$ and/or $H_2D^+$ emission observations available (Phillips et al. 1992; van Dishoeck et al. 1992), as well as infrared absorption line data on $H_3^+$ (Black et al. 1990) and $CO/H_2$ (Lacy et al. 1994). Previous studies in cold dark clouds such as TMC-1, with $n(H_2) = 10^4$ cm$^{-3}$, have found $5.8\ 10^{-9} \leq x_e \leq 4.9\ 10^{-8}$, depending on the amount of PAHs and metals included in the modeling (Schilke et al. 1991). The sources studied in this work have higher temperatures and densities up to two orders of magnitude, and it will be interesting to investigate whether the electron fraction in them is lower, as predicted by theory. The observations and results are presented in Sections 2–3, and their analysis in Section 4.

Together, the abundances of these ions provide a firm lower limit to the ionization fraction of the cloud. In order to constrain the electron abundance, more detailed modeling is needed. The basic idea is that the molecular ions are formed by protonation of abundant neutral species in reactions with $H_3^+$:

$$X + H_3^+ \rightarrow XH^+ + H_2$$

where X=CO, $H_2O$, $N_2$, CS... Their destruction is governed by reactions with CO, $H_2O$ and O and by dissociative recombination with an electron in the gas phase:

$$XH^+ + e \rightarrow X + H$$

Thus by measuring $XH^+$ and the abundances of the parents X for a sufficient number of species, it should be possible to constrain both the electron fraction and the $H_3^+$ abundance (or equivalently the cosmic ray ionization rate $\zeta_H$) in dense clouds. This technique was pioneered by Wootten et al. (1979) based on the $HCO^+$/CO ratio, but is extended here to include many more ion/neutral pairs. The actual modeling, which involves a more detailed network of reactions, is presented in Section 5.

## 2 Observations

The sources studied in this work are both associated with active regions of star formation. NGC 2264 IR is a well studied molecular cloud core ($d \sim 800$ pc) (Schwartz et al. 1985; Krügel et al 1987). Its shape on an optical plate resembles a dark cone (the "Cone Nebula"). Presumably an intermediate mass star is forming here. The infrared source is bright (Allen 1972) and has been used as background light source for infrared absorption line studies of gas and solid phase species. In particular, an upper limit of $1.7 \, 10^{-8}$ exists on the $H_3^+$ abundance in the gas (Black et al. 1990), and a lower limit of $1.7 \, 10^{-4}$ on the $CO/H_2$ ratio (Lacy et al. 1994).

W 3 IRS5 is the brightest infrared source ($L = 1.7 \, 10^5$ $L_\odot$ (Ladd et al. 1993)) in the W 3 Giant Molecular Cloud core ($d \sim 2.3$ kpc). A large amount of dense and warm material surrounds the massive star forming inside (Dickel 1980; Dickel et al. 1980; Hasegawa et al. 1994; Hayashi et al. 1989; Tieftrunk et al. 1995; Helmich et al. 1994). Because of the enormous visual extinction ($A_V > 100$ mag) the ultraviolet radiation from the young stars is not expected to influence the ionization balance deep inside the cloud.

The observations of NGC 2264 were performed primarily at the James Clerk Maxwell

Telescope (JCMT[1]) in February 1994, with additional data taken in April, June and November 1994. The facility receivers at 230 GHz (A2), 345 GHz (B3i) and 460 GHz (C2) were employed. These are double side-band receivers with upper and lower side-bands 3.00 GHz (A2, B3i) or 7.88 GHz (C2) apart. No systematic side-band checks were performed, since the spectra are simple and often there is only a single line in the spectrum. The image frequencies were compared against spectral-line catalogues and no obvious blending of lines was found. As the backend, the Digital Autocorrelation Spectrometer (DAS) was used in different spectral resolutions (mostly 0.094, 0.187 and 0.374 MHz/channel). To improve the signal to noise, the spectra were smoothed to 0.374 MHz/channel and in some cases to 0.750 MHz/channel. The former corresponds to $\sim$0.5 km s$^{-1}$ at 230 GHz. Integration times were generally 30 minutes on+off for 230 and 345 GHz receivers but up to an hour for lines in the 460 GHz window, because of the higher system temperatures. This resulted in typical 1$\sigma$ noise in $T_A^*$ of 30 (A2)–50 (B3i) mK and 200–350 (C2) mK per resolution element (0.374 MHz). The pointing was checked regularly on OMC-1 and was found to be within 3$''$.

The calibration at the JCMT was performed with the chopper-wheel method. In February 1994, the beam efficiency was somewhat uncertain, because of inaccuracies in the shape of the telescope surface. However, careful comparison with earlier runs showed a reduction by only $\sim$10%. This effect is corrected for by using an effective main beam-efficiency of 0.53, 0.45 and 0.35 for receiver A2, B3i and C2, respectively. During other runs, values of 0.72, 0.60 and 0.42, respectively, were used. In general we expect the absolute calibration to be accurate to about 30%; relative uncertainties can be significantly smaller for all runs. The beam size of the JCMT at 230, 345 and 460 GHz is $\sim$20$''$, 15$''$ and 12$''$, respectively. For most spectra, position-switching by 600$''$ was found to be appropriate, except for the $^{12}$CO lines for which a larger switch was used. Additional data on other species toward NGC 2264 IRS1 will be presented by Schreyer et al. (1996).

A limited number of spectra of NGC 2264 were taken at the Caltech Submillimeter Observatory (CSO[2]) in January 1994 with the 230 GHz (H$_2$CO at 218 GHz) and 345 GHz (C$^{17}$O 3–2) receivers. The low (500 MHz) and high (50 MHz) resolution acousto-optical spectrometers were used as backends. Pointing is reproducible within 5$''$ on the CSO. Main beam efficiencies of 0.72 and 0.60 were adopted, respectively. The CSO data at 218 GHz refer to a beam of $\sim$32$''$; those at 345 GHz to a 20$''$ beam.

---

[1] The James Clerk Maxwell Telescope is operated by the Observatories on behalf of the Particle Physics and Astronomy Research Council of the United Kingdom, the Netherlands Organisation for Scientific Research, and the National Research Council of Canada.

[2] The Caltech Submillimeter Observatory is operated by the Californian Institute of Technology under funding from the U.S. National Science Foundation (AST 93-13929).

Earlier observations by van Dishoeck et al. (1992) with the CSO have resulted in upper limits on the $H_2D^+$ abundance. These same spectra also reveal surprisingly strong $N_2H^+$ 4–3 emission in this source in a 18″ beam.

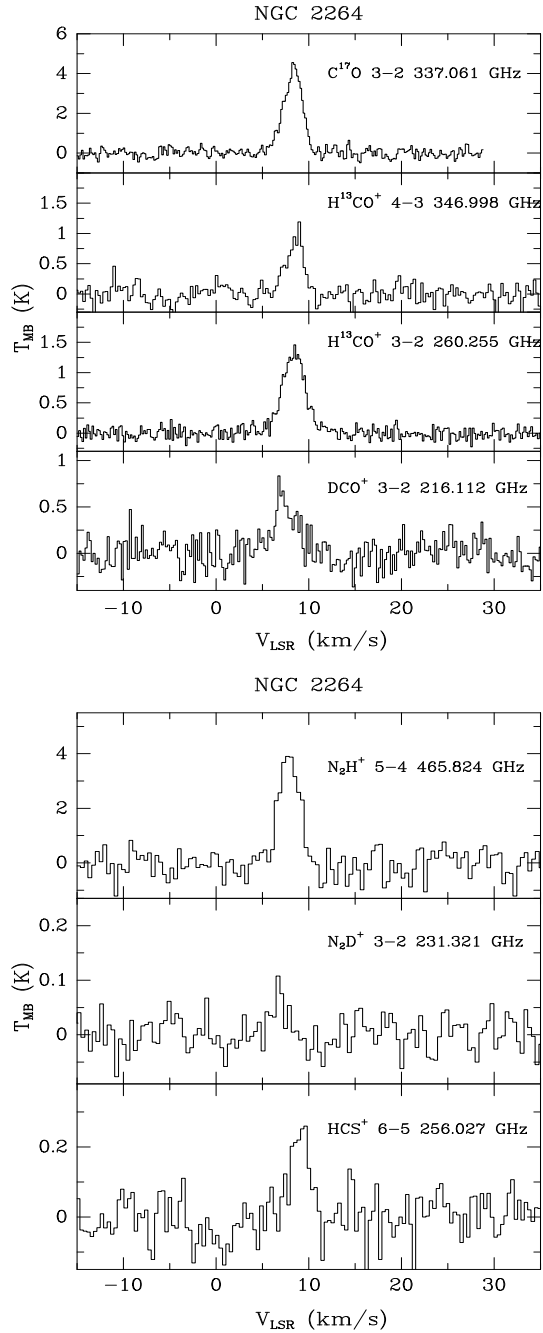

The observations presented here for W 3 IRS5 form part of the large JCMT 345 GHz survey of three star-forming cores in the W 3 cloud (IRS4, IRS5 and W 3($H_2O$)) between January 1992 and November 1994. A complete log and details of the observations will be given in a separate publication concerning the spectral survey (Helmich & van Dishoeck 1995; see also Helmich et al. 1994 for first results). For all runs, the facility receivers (A1, A2, B3i and C2) were used. Prior to July 1993, the acousto-optical spectrometer AOSC (500 MHz) was employed as the backend; subsequently the DAS was used in different spectral resolutions, mostly in its 500 MHz bandwidth mode. Binning of two channels resulted in a typical resolution of 0.750 MHz/channel. As for NGC 2264, no systematic side band identification was performed, because the line crowding in IRS5, compared with the two other W 3 sources in the spectral-line survey, is such that there is little ambiguity and possible blends are easily identified. The pointing was checked regularly on the nearby continuum source W 3(OH) and is within 3″. The observations of W 3 IRS5 were performed in the 180″ beam-switch mode, except for the $^{12}$CO lines where a larger position switch was used.

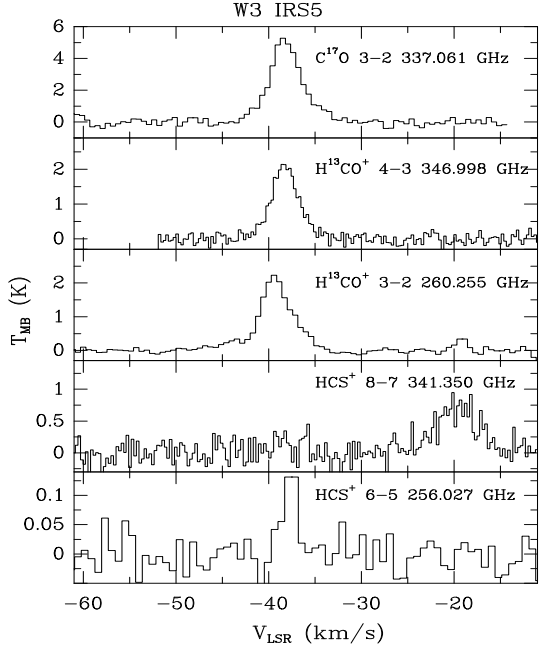

Figure 1: Observed spectra toward: **a and b** NGC 2264 IRS1 and **c** W 3 IRS5. Note that the DCO$^+$ 3–2 line toward NGC 2264 consists of two components. Only the broad component was used in the analysis. The N$_2$H$^+$ 5–4 line represents the first detection of this line in the interstellar medium. The N$_2$D$^+$ 3–2 line may be present at the 2$\sigma$ level. Since W 3 IRS5 is a well-known outflow source, only the central Gaussian was used in the analysis. The line next to the HCS$^+$ 8–7 feature at $V_{\rm LSR} = -20$ km s$^{-1}$ is the $^{34}$SO$_2$ $19_{1,19} - 18_{0,18}$ line from the image side-band.

Additional observations of the $H_3O^+$ ion toward W 3 IRS5 were obtained by Phillips et al. (1992) using the CSO, whereas an upper limit on $H_2D^+$ was obtained by van Dishoeck et al. (1992). The latter study also reported detection of the $N_2H^+$ 4–3 line.

## 3 Results

Tables 1 and 2 summarize the resulting Gaussian fit parameters for the various lines for the two sources. Examples of spectra are given in Fig. 1 a, b and c. Although the profiles of the lower lines of $DCO^+$ and other species toward NGC 2264 suggest the presence of two components (Guélin et al. 1982; Schreyer et al. 1996), it was found that the lines observed in this work are well represented by single Gaussians with little spread in the line widths, the optically thick $^{12}$CO 4–3 and $HCO^+$ 4–3 lines being exceptions. Most other lines are optically thin, or at most moderately optically thick. For example, the $C^{34}S$ 5–4 line is weak, and the intensity ratio $CS/C^{34}S$=16–21 is only slightly smaller than the cosmic $[^{32}S]/[^{34}S]$ abundance ratio of 22.59. As mentioned above, $N_2H^+$ is very strong in this source; the 5–4 line at 465 GHz presented in Fig. 1b is the first detection of this line in any astrophysical source. $N_2D^+$ 3–2 appears present as well, but only at the $2\sigma$ level.

The lines toward W 3 IRS5 are also well represented by single Gaussians but there is a much larger spread in the line widths. The molecules $HCS^+$, $C^{33}S$ and $C^{34}S$ have much narrower lines than e.g., $H^{13}CO^+$, $HCO^+$ and CO-isotopomers. It is easy to explain the larger line widths for CO and $HCO^+$ by optical depth effects, but this becomes more difficult for the rarer species. Part of the explanation may lie in inaccurate

Table 1: Parameters of Gaussian fits for NGC 2264

| Line | $T_{\rm MB}$ (K) | $\Delta V$ km s$^{-1}$ | $\int T_{\rm MB} dV$ K km s$^{-1}$ |
|---|---|---|---|
| CO 4 − 3 | 54.3 | 10.1 | 586 |
| C$^{18}$O 2 − 1 | 6.74 | 3.0 | 21.1 |
| C$^{17}$O 3 − 2 | 4.18 | 2.3 | 10.3 |
| CS 5 − 4 | 8.48 | 3.3 | 30 |
| CS 7 − 6 | 5.42 | 3.1 | 18 |
| CS 10 − 9 | ≤ 2.44 | | |
| C$^{34}$S 5 − 4 | 0.53 | 2.5 | 1.42 |
| HDO $1_{01} - 0_{00}$ | ≤ 1.94 | | |
| HCO$^+$ 4 − 3 | 13.6 | 4.1 | 59.3 |
| H$^{13}$CO$^+$ 3 − 2 | 1.34 | 2.8 | 4.0 |
| H$^{13}$CO$^+$ 4 − 3 | 0.96 | 2.1 | 2.1 |
| DCO$^+$ 3 − 2 | 0.42 | 3.2 | 1.4 |
| HCS$^+$ 6 − 5 | 0.26 | 2.2 | 0.64 |
| N$_2$H$^+$ 5 − 4 | 4.1 | 2.5 | 10.9 |
| N$_2$D$^+$ 3 − 2 | 0.11 | 0.8 | 0.08 |
| [C I]$^3$P$_1$ − $^3$P$_0$ | 20.6 | 4.5 | 98.9 |

Coordinates (B1950.0): $\alpha$ = 06 38 25.0, $\delta$ = 09 32 29.
All data were taken with the JCMT except for C$^{17}$O 3 − 2.
The upper limits are 2$\sigma$ in a 0.625 MHz channel.

determinations of the C$^{33}$S and HCS$^+$ line widths, because these lines are so weak. However, source structure can play a rôle as well. A complete investigation is beyond the scope of this paper; therefore, for each species its observed line width is adopted in the analysis. The detection of C$^{33}$S implies that even the C$^{34}$S lines are slightly optically thick. From the $^{13}$CS line, however, it can be seen that the optical depth must be small.

Table 2: Parameters of Gaussian fits for W 3 IRS5

| Line | $T_{MB}$ (K) | $\Delta V$ km s$^{-1}$ | $\int T_{MB} dV$ K km s$^{-1}$ |
|---|---|---|---|
| CO 3−2 | 77.7 | 13.2 | 1099 |
| CO 2−1 | 103 | 10.7 | 1177 |
| C$^{18}$O 2 − 1 | 11.2 | 4.59 | 54.7 |
| C$^{17}$O 3 − 2 | 3.87 | 4.09 | 16.9 |
| CS 7 − 6 | 6.53 | 3.66 | 25.4 |
| C$^{33}$S 5 − 4 | 0.19 | 2.84 | 0.58 |
| C$^{33}$S 7 − 6 | 0.59 | 1.35 | 0.85 |
| C$^{34}$S 5 − 4 | 1.14 | 2.91 | 3.54 |
| C$^{34}$S 7 − 6 | 0.77 | 1.83 | 1.5 |
| $^{13}$CS 5 − 4 | 0.35 | 2.77 | 1.04 |
| HCO$^+$ 4 − 3 | 24.8 | 6.15 | 162 |
| H$^{13}$CO$^+$ 3 − 2 | 2.06 | 4.21 | 9.28 |
| H$^{13}$CO$^+$ 4 − 3 | 2.49 | 3.47 | 9.18 |
| HC$^{18}$O$^+$ 4 − 3 | 0.15 | 3.3 | 0.56 |
| DCO$^+$ 5 − 4 | 0.22 | 2.67 | 0.64 |
| HCS$^+$ 6 − 5 | 0.23 | 0.88 | 0.21 |
| HCS$^+$ 8 − 7 | 0.28 | 1.58 | 0.46 |
| N$_2$H$^+$ 5 − 4 | <0.38 | | |
| N$_2$D$^+$ 3 − 2 | < 0.1 | | |
| [C I]$^3$P$_1$ −$^3$ P$_0$ | 16.5 | 7.05 | 124 |

Coordinates (B1950.0): $\alpha$ = 02 21 53.1, $\delta$ = 61 52 20.

All data were taken with the JCMT.

The upper limits are 2$\sigma$ in a 0.625 MHz channel.

# 4 Analysis

## 4.1 Physical parameters

In order to derive the column densities from the observed line intensities, first the physical parameters in the sources have to be determined. Statistical equilibrium calculations were performed along the lines described by Jansen et al. (1994) and Jansen (1995). These models take the various collisional and radiative excitation and de-excitation processes into account, and treat the radiative transfer with an escape probability method. Because the observed lines are mostly optically thin, the precise details of the radiative transfer method do not affect the results. The ratios of some $H_2CO$ lines are particularly sensitive to temperature variations (see e.g. Jansen et al. 1993; Helmich et al. 1994; Mangum & Wootten 1993) and are thus very well suited for kinetic temperature determinations. Lines from linear rotors such as CS, $C^{34}S$, or ions like $H^{13}CO^+$, on the other hand, are more sensitive to density.

Line intensity ratios were calculated for a range of densities and temperatures for $H_2CO$, CS, $H^{13}CO^+$ and $N_2H^+$, and are plotted in Fig. 2a, together with the observed ratios for NGC 2264 (see also Schreyer et al. 1996). The contour plot of the CS 7–6/5–4 line ratio is produced for the slightly optically thick case appropriate for this source. All other ratios refer to optically thin lines. From the $H_2CO$ $3_{03}-2_{02}$ and $3_{22}-2_{21}$ lines at 218 GHz measured in the same frequency setting, a kinetic temperature of $55^{+10}_{-5}$ K is found. Furthermore it is seen from Fig. 2a that the measurements for the $N_2H^+$ and CS ratios indicate a similar density of $(2-3)\,10^6$ cm$^{-3}$. Thus, these species probably probe the same region. The density indicated by $H^{13}CO^+$ appears to be somewhat lower than that found from the two other species. This difference is difficult to account for in terms of optical thickness, but could be due, e.g., to uncertainties in the collisional rate coefficients or source structure. The measurements are, however, within the errors in agreement with each other and give a number density of $n(H_2)=(2\pm1.5)\,10^6$ cm$^{-3}$. Maps of the CS 7–6 and 5–4 and $^{12}CO$ 3–2 lines by Schreyer et al. (1996) show that the temperature cannot be much lower than that found from $H_2CO$. Note that the above single temperature and density analysis assumes that all species and lines are similarly distributed in the beam. No corrections are made for the fact that the higher frequency JCMT data refer to a smaller beam than those at lower frequencies.

For W 3 IRS5 the temperature $T_{\rm kin} \approx 100^{+40}_{-20}$ K and density $n(H_2) \approx 10^6$ cm$^{-3}$ derived in Helmich et al. (1994) from various formaldehyde lines was adopted. The $H^{13}CO^+$ 4–3/3–2 and $C^{34}S$ 7–6/5–4 line ratios, presented in Fig. 2b, show that these molecules probe about the same density, suggesting that they are present in the same gas. It is assumed that the emission of all molecules described here fills our beams.

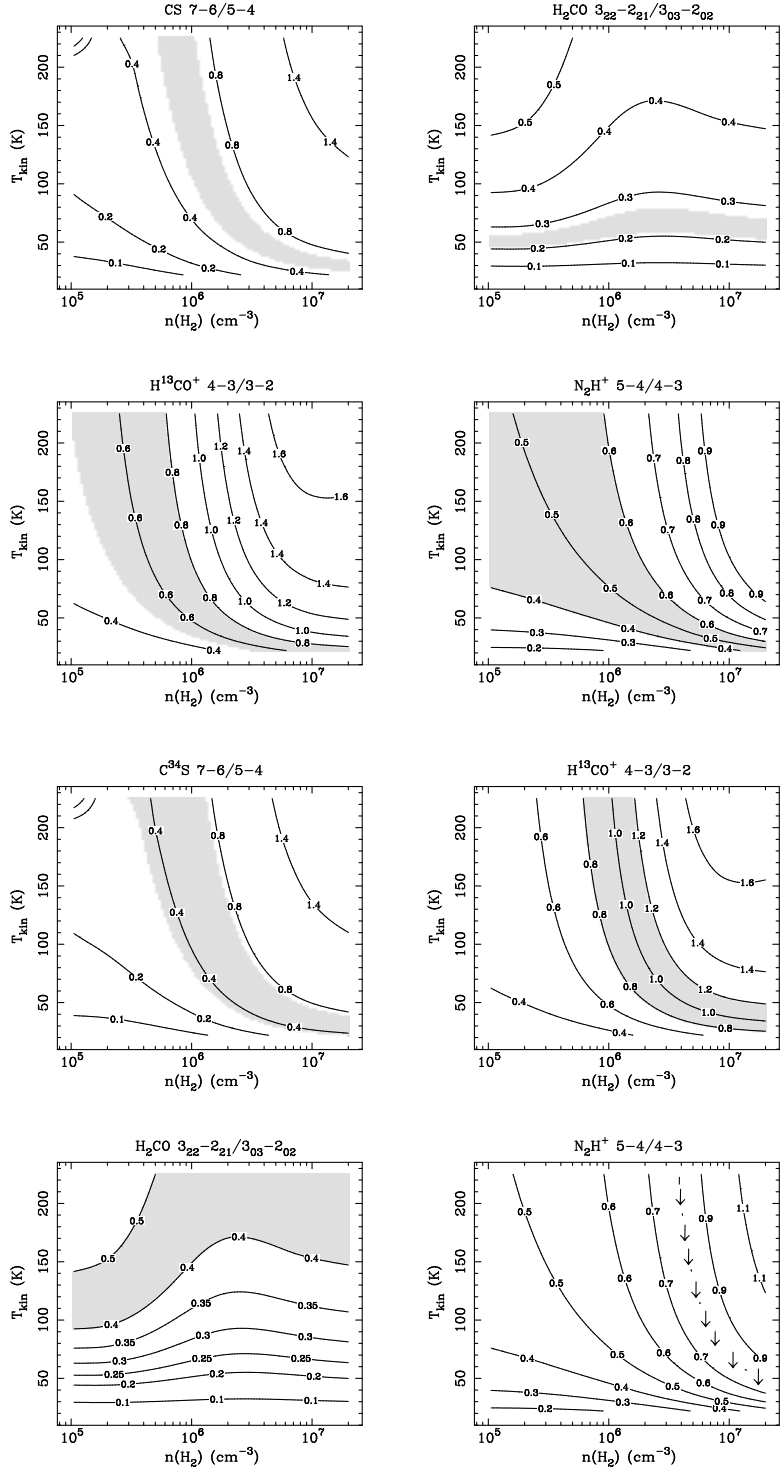

Figure 2: Contour plot showing ratios of lines. Shaded regions refer to the observed range toward the source: **a** NGC 2264, **b** W 3 IRS5. All ratios refer to optically thin lines, except the CS 7–6/5–4 ratio toward NGC 2264, where $N(\mathrm{CS}) = 1.3\,10^{14}$ cm$^{-2}$ and $\Delta V = 3.2$ km s$^{-1}$ was adopted. The N$_2$H$^+$ ratio for W 3 IRS5 is uncertain and is therefore plotted as an upper limit.

The inferred densities toward our two sources are indeed two orders of magnitude larger than those found in dark clouds such as TMC-1, which have been the subject of previous studies of the electron fraction.

## 4.2 Column densities

The total beam-averaged column densities for each species are calculated by fitting the absolute intensities, using the temperature and density derived from our statistical equilibrium calculations. The results are given in Tables 3 and 4 for the two sources. Since the HCO$^+$ 4–3 line is optically thick toward NGC 2264, the column density of HCO$^+$ is determined from H$^{13}$CO$^+$, using the [$^{12}$C/$^{13}$C] ratio of 94 obtained by Mitchell & Maillard (1993) for this source. The $^{12}$CO column density was calculated from the C$^{17}$O 3–2 line assuming that the CO/C$^{17}$O ratio is the same as the cosmic [$^{16}$O]/[$^{17}$O] ratio of 2600. The lower limit on the CO/H$_2$ ratio found by Lacy et al. (1994) of 1.7 10$^{-4}$ then gives an upper limit on the total beam-averaged column-density $N$(H$_2$)$\approx$ 8.3 10$^{22}$ cm$^{-2}$. Although the CO/H$_2$ ratio derived from the pencil-beam infrared absorption line observations does not necessarily apply to the whole volume sampled by the emission data, the CO column density derived from the two methods agree within a factor of two.

The total (beam-averaged) column density toward W 3 IRS5 is derived from the C$^{17}$O 3–2 line. In contrast with Helmich et al. (1994), a CO/H$_2$ abundance ratio of 2.7 10$^{-4}$ was used instead of 8 10$^{-5}$. The value in Helmich et al. (1994) was chosen because the study of Tielens et al. (1991) of the solid CO seemed to indicate a lower value. The subsequent study of Lacy et al. (1994) however suggests a much larger ratio for warm sources like NGC 2024 IRS2. Since the gas and grain temperatures toward W 3 IRS5 are even higher than those toward NGC 2024, a higher value is probably more appropriate. This value decreases the total beam-averaged column density from 5 10$^{23}$ cm$^{-2}$ to 1.3 10$^{23}$ cm$^{-2}$, and at the same time increases the abundances by a similar factor. It is stressed again, however, that the derived column-densities and abundances are uncertain for this source due to the unknown structure and distribution of the various species within the beam. The relative values should be more reliable, however.

For both clouds the observation of the [C I] 492 GHz line provides good upper limits on the beam averaged column density and abundance of atomic carbon since a large fraction of this emission probably arises from the outer skin of the cloud.

Tables 3 and 4 include column densities and abundances (w.r.t. H$_2$) for other relevant species obtained from the literature for the two sources. In the case of NGC 2264 we adopted the H$_2$D$^+$ column density derived by van Dishoeck et al. (1992), with slightly different total H$_2$ density and temperature. The H$_2$O column density is taken from Wannier et al. (1991) who derived H$_2$O/CO $< 0.003$ in NGC 2264 from searches for H$_2^{18}$O using the Kuiper Airborne Observatory in a 2.7$'$ beam. For H$_3$O$^+$ toward W 3 IRS5, the data given in Phillips et al. (1992) were reanalysed with the above density and temperature. From observations of the H$_2$O 183 GHz line presented by Cernicharo et al. (1990), Phillips et al. (1992) determined $x$(H$_2$O) $\sim 10^{-6} - 10^{-5}$. The O$_2$ limit of $< 10^{19}$ cm$^{-2}$ for that source is based on a deep search for the $^{16}$O$^{18}$O 234 GHz line (Keene, van Dishoeck & Phillips, unpublished result).

Table 3: Column densities and abundances in NGC 2264

| Species | $N$ (cm$^{-2}$) | $x$ |
|---|---|---|
| CS | $1.3\ 10^{14}$ | $1.6\ 10^{-9}$ |
| C$^{34}$S | $4.0\ 10^{12}$ | $4.8\ 10^{-11}$ |
| HDO | $\leq 7.1\ 10^{13}$ | $\leq 8.6\ 10^{-10}$ |
| H$^{13}$CO$^+$ | $1.6\ 10^{12}$ | $1.9\ 10^{-11}$ |
| HCO$^+$ | $1.5\ 10^{14}$ | $1.8\ 10^{-9}$ |
| DCO$^+$ | $7.4\ 10^{11}$ | $8.9\ 10^{-12}$ |
| HCS$^+$ | $1.8\ 10^{12}$ | $2.2\ 10^{-11}$ |
| N$_2$H$^+$ | $2.5\ 10^{13}$ | $3.0\ 10^{-10}$ |
| N$_2$D$^+$ | $4.6\ 10^{10}$ | $5.5\ 10^{-13}$ |
| H$_2$D$^+$ | $< 9.0\ 10^{12}$ | $< 1.0\ 10^{-10}$ |
| H$_3^+$ | $< 1.4\ 10^{15}$ | $< 1.6\ 10^{-8}$ |
| H$_2$O | $< 4.2\ 10^{16}$ | $< 5.1\ 10^{-7}$ |
| C | $\leq 1.8\ 10^{18}$ | $\leq 2.2\ 10^{-5}$ |

# 5  Models

In order to estimate the electron fraction, we built a simple chemical model which calculates the steady state abundances of the major positive ions, namely: H$^+$, H$_2^+$, H$_3^+$, He$^+$, C$^+$, H$_3$O$^+$, HCO$^+$, DCO$^+$, N$_2$H$^+$, H$_2$D$^+$, HCS$^+$, N$_2$D$^+$ and H$_2$DO$^+$. The electron density is set equal to the sum of the densities of the positive ions, including metals or PAH ions. Only gas phase reactions are included. The temperature, density of molecular hydrogen and abundances of CO and CS are those derived from the observations for each cloud.

Table 4: Column densities and abundances in W 3 IRS5

| Species | $N$ (cm$^{-2}$) | $x$ |
|---|---|---|
| CS | $2.0\ 10^{14}$ | $1.5\ 10^{-9}$ |
| C$^{17}$O | $1.4\ 10^{16}$ | $1.1\ 10^{-7}$ |
| H$^{13}$CO$^+$ | $3.3\ 10^{12}$ | $2.5\ 10^{-11}$ |
| HCO$^+$ | $2.0\ 10^{14}$ | $1.5\ 10^{-9}$ |
| DCO$^+$ | $4.8\ 10^{11}$ | $3.7\ 10^{-12}$ |
| HCS$^+$ | $6.3\ 10^{11}$ | $4.8\ 10^{-12}$ |
| N$_2$H$^+$ | $\leq 8.3\ 10^{11}$ | $\leq 6.4\ 10^{-12}$ |
| N$_2$D$^+$ | $\leq 3.5\ 10^{11}$ | $\leq 2.3\ 10^{-12}$ |
| H$_3$O$^+$ | $5.6\ 10^{13}$ | $4.3\ 10^{-10}$ |
| H$_2$D$^+$ | $\leq 5.8\ 10^{12}$ | $\leq 4.4\ 10^{-11}$ |
| O$_2$ | $\leq 1.0\ 10^{19}$ | $\leq 7.7\ 10^{-5}$ |
| C | $\leq 2.1\ 10^{18}$ | $\leq 1.6\ 10^{-5}$ |

The abundances of O, O$_2$, H$_2$O, N, N$_2$, D and HD are free parameters, taking into account the constraints imposed by other observations or models, and allowing for a certain amount of depletion on grains. The main motivation of this approach is that detailed chemical models generally give O$_2$ and H$_2$O abundances higher than the upper limits derived from the observations ($x \sim 10^{-6} - 10^{-5}$) (see Tables 3 and 4). Moreover these models usually do not allow for grain surface reactions and do not consider deuterated varieties. In our case the abundance of O$_2$ does not play an important role since this species contributes solely to the destruction of He$^+$, H$^+$ and C$^+$, for which we do not have observed abundances, and for which the calculated abundances are low in any case. Other free parameters are the cosmic ray ionization rate $\zeta_H$ and the abundance of ionized metal $x_{M^+}$.

Concerning the reaction rates, special attention has been paid to the dissociative recombination rate coefficients of HCS$^+$, HCO$^+$, N$_2$H$^+$ and H$_3^+$, which have recently been studied in the lab. For HCO$^+$ and H$_3^+$, the rate coefficients measured by Rowe et al. (1993) and Amano (1990) are very close: from their work, we derive the temperature-dependent values:

$$k_{\text{HCO+}} = 2.9\ 10^{-7}(T/300)^{-0.71}\ \text{cm}^3\text{s}^{-1},$$

$$k_{\text{H}_3^+} = 1.8\ 10^{-7}(T/300)^{-0.88}\ \text{cm}^3\text{s}^{-1}.$$

In fact, Smith & Španel (1993) obtain a lower value $k_{\text{H}_3^+} = 3\ 10^{-8} \text{cm}^3\text{s}^{-1}$ at 300 K, but use of this value does not affect our results since $\text{H}_3^+$ is mainly destroyed by CO, due to the high density and the high abundance of CO.

For $\text{N}_2\text{H}^+$, the rate measured by Amano (1990) can be fitted by:

$$k_{\text{N}_2\text{H+}} = 7.3\ 10^{-7}(T/300)^{-0.59}\ \text{cm}^3\text{s}^{-1}.$$

For $\text{HCS}^+$, Rowe et al. (1993) found a rate coefficient of $(6 - 8)\ 10^{-7}\ \text{cm}^3\text{s}^{-1}$ at 300 K, which is about one order of magnitude larger than found in previous studies (Millar et al. 1985; Talbi et al. 1989). The original motivation for slow $\text{HCS}^+$ dissociative recombination came from the observed high $\text{HCS}^+/\text{CS}$ abundance ratios. Rowe et al. (1993) claim that they can still reproduce this high value with a rapid rate by assuming that the branching ratio toward the H + CS channel represents only one third of the total dissociative recombination process. Since, in addition to this assumption, the influence of the vibrational excitation of the $\text{HCS}^+$ ions in their experiment is unclear, the original determination of Millar et al. is kept:

$$k_{\text{HCS+}} = 5\ 10^{-8}(T/300)^{-0.5}\ \text{cm}^3\text{s}^{-1}$$

For the deuterated species ($\text{DCO}^+$, $\text{H}_2\text{D}^+$, $\text{N}_2\text{D}^+$ and $\text{H}_2\text{DO}^+$) the same dissociative recombination rate coefficients as for the hydrogen-based ones have been adopted; the main formation and destruction rate coefficients are those given by the UMIST group (Millar et al. 1991). In particular, we followed the study of Millar et al. (1989) for the temperature-dependence of the important exchange reaction, $\text{H}_3^+ + \text{HD} \rightarrow \text{H}_2\text{D}^+ + \text{H}_2$:

$$k_{\text{f}} = 1.8\ 10^{-9}(1. - T/340)\ \text{cm}^3\text{s}^{-1}$$

For the reverse reaction, Pagani et al. (1992) have shown that the rate depends on the ortho-$\text{H}_2\text{D}^+$/para-$\text{H}_2\text{D}^+$ ratio. Since it is beyond the scope of this study to calculate this ratio in our model, we simply adopt: $k_{\text{r}} = k_{\text{f}} e^{-227/T}$.

The best fitting models for the two clouds are described in the next section. The model parameters are summarized in Tables 5 and 7.

## 5.1 NGC 2264 IRS1

The principal characteristics of the NGC 2264 observations are the high abundance of $N_2H^+$, and the high $HCS^+/CS$ ratio equal to $1.4\ 10^{-2}$. As noted by van Dishoeck et al. (1992), the high $N_2H^+$ abundance implies that virtually all nitrogen must be in the gas phase in the form of $N_2$. Also, the $CO/H_2$ ratio has to be close to the lower limit derived by Lacy et al. (1994), i.e., $CO/H_2 = 1.7\ 10^{-4}$, because the dominant destruction route for $N_2H^+$ is by reaction with CO, and the main formation route is by: $N_2 + H_3^+ \rightarrow N_2H^+ + H_2$, and $H_3^+$ is also mainly destroyed by CO. $H_3^+$, in turn, is formed through cosmic ray ionization of $H_2$, and its abundance is directly proportional to $\zeta_H$ (Lepp et al. 1987; Black et al. 1990; Phillips et al. 1992).

As can be seen from Tables 5 and 6, the best fit requires that all nitrogen is in gas phase $N_2$ for a rather high value of $\zeta_H \approx 2.5\ 10^{-16}$ s$^{-1}$ and low abundances of atomic oxygen and carbon (Model 1). In fact, if $x(O) \ll 8.5\ 10^{-4}$ and $x(C) \ll 1.4\ 10^{-4}$, the abundance of $N_2H^+$ can be written as:

$$\left[\frac{x(N_2H^+)}{3\ 10^{-10}}\right] \sim \frac{\left[\frac{x(N_2)}{1.1\ 10^{-4}}\right]\left[\frac{\zeta_H}{1.4\ 10^{-16}}\right]}{\left(1 + \frac{x(H_2O)}{5.7\ 10^{-5}} + \frac{x_e}{7.4\ 10^{-8}}\right)\left(1 + \frac{x_e}{3.6\ 10^{-7}}\right)}$$

Thus, the $N_2H^+$ abundance can also be reproduced by an even higher value of $\zeta_H$, but with the fraction of nitrogen in $N_2$ decreased. The dependence of the ion abundances on $\zeta_H$ is illustrated in Fig. 3a. The inferred values for $\zeta_H$ are larger than those normally assumed for dense clouds, $\zeta_H \approx (1-5)\ 10^{-17}$ s$^{-1}$, but could be due to e.g., the presence of X-rays from the young stellar object, which have a similar effect on the ionization (Casanova et al. 1995, Farquhar et al. 1994). If the abundances of O and C are close to the upper limits derived from observations and other estimates (Little et al. 1994, 1995; Pagani & Encrenaz 1994) $\zeta_H$ has to be very high: an extreme case is given by Model 2, in which we have assumed $x(O) = x(CO)$ and $x(C) = 0.1\ x(CO)$.

The main problem comes from reproducing the high $HCS^+$ abundance. It can be shown that its abundance can be expressed as:

$$\left[\frac{x(HCS^+)}{2.2\ 10^{-11}}\right] \sim 0.3 \frac{x(H_3O^+) + 1.2x(HCO^+) + 17.2x(H_3^+)}{x_e + 5.6\ 10^{-5}x(O)}$$

Figure 3: Ratios of calculated ion abundances to observed ones as functions of the three dominant parameters: $\zeta_H$, x($H_2O$) and x($M^+$): **a** in the case of NGC 2264 (Model 1), **b** in the case of W 3 IRS5 (Model 2).

Table 5: Model inputs for NGC 2264

| Parameter | Model 1 | Model 2 |
|---|---|---|
| $\zeta_H$ (s$^{-1}$) | $2.5\ 10^{-16}$ | $3.\ 10^{-15}$ |
| $x(M^+)$ | $\leq 2\ 10^{-9}$ | $\leq 10^{-9}$ |
| $x(N_2)$ | $1.1\ 10^{-4}$ | $3.1\ 10^{-5}$ |
| $x(H_2O)$ | $3\ 10^{-5}$ | $10^{-4}$ |
| $x(O)$ | $\leq 5.\ 10^{-6}$ | $1.7\ 10^{-4}$ |
| $x(C)$ | $\leq 5.\ 10^{-6}$ | $1.7\ 10^{-5}$ |
| $x(D)$ | $2.8\ 10^{-7}$ | $8.\ 10^{-7}$ |
| $x(HD)$ | $2.8\ 10^{-5}$ | $10^{-5}$ |

so that $x(HCS^+)$ increases when the electron fraction decreases: it is one of the reasons why the metal ion abundance has to be low in this case. Moreover, keeping in mind that the abundance of $H_3^+$ is now fixed – because $\zeta_H$ is fixed from the $N_2H^+$ observations – , $x(HCS^+)$ depends essentially on the abundances of $HCO^+$ and $H_3O^+$, where $x(HCO^+)$ has been constrained from observations. As can be seen in Fig. 3a, the dependence of the $HCS^+$ abundance on $\zeta_H$ is not large, because in the model $H_3O^+$ is the dominant ion leading to the formation of $HCS^+$.

The formation of $HCO^+$ invokes CO, $H_3^+$ and $N_2H^+$. Its destruction proceeds mainly by dissociative recombination and by reaction with $H_2O$ and atomic deuterium, which results in $DCO^+$. $H_3O^+$ is formed by reactions of $H_2O$ and O with molecular ions (essentially $HCO^+$ and $H_3^+$) and is destroyed by dissociative recombination.

At this point, the important remaining parameters are thus the abundances of $H_2O$, D and HD which govern the abundances of $HCO^+$, $H_3O^+$ and $DCO^+$, the latter via $HCO^+$ and $H_2D^+$. Fig. 3a shows the model $HCO^+$ and $DCO^+$ abundances as functions of the $H_2O$ abundance. In order to reproduce the observed values a rather high abundance of $H_2O$ of at least $3\ 10^{-5}$ is needed. Also, it is inferred that only 1 to 5% of deuterium is in atomic form. The high value of $x(H_2O)$ seems in contradiction with the work of Wannier et al. (1991). However, their results refer to a much larger beam of $\sim 2.7'$ than the region sampled by our observations. The presence of water masers near IRS1 suggests that locally the $H_2O$ abundance may be significantly higher (Henning et al. 1992).

Finally, as can be seen in Table 6, good agreement between the observed and calculated abundances is achieved for $N_2H^+$, $HCO^+$, $DCO^+$, $N_2D^+$ and $HCS^+$.

Table 6: Model fits for NGC 2264

| $x(\text{model})/x(\text{obs})$ | Model 1 | Model 2 |
|---|---|---|
| HCO$^+$ | 1.0 | 1.7 |
| DCO$^+$ | 0.8 | 1.1 |
| N$_2$H$^+$ | 1.0 | 1.0 |
| N$_2$D$^+$ | 1.4 | 2.4 |
| HCS$^+$ | 0.7 | 0.8 |
| predictions | | |
| $x_e$ | 9.6 10$^{-9}$ | 3.3 10$^{-8}$ |
| $x(\text{H}_3^+)$ | 3.9 10$^{-10}$ | 2.9 10$^{-9}$ |
| $x(\text{H}_2\text{D}^+)$ | 6.7 10$^{-13}$ | 1.9 10$^{-12}$ |
| $x(\text{H}_3\text{O}^+)$ | 6.8 10$^{-9}$ | 2.5 10$^{-8}$ |
| $x(\text{H}_2\text{DO}^+)$ | 2.0 10$^{-11}$ | 3.1 10$^{-11}$ |
| $x(\text{NH}_3)$ | 5.1 10$^{-9}$ | 3.1 10$^{-9}$ |
| $x(\text{NH}_2\text{D})$ | 1.9 10$^{-12}$ | 5.0 10$^{-13}$ |

With these assumptions, the electron fraction is $x_e = (1.0 - 3.3)\ 10^{-8}$, the dominant ion being H$_3$O$^+$. Searches for this ion in this object would therefore be very valuable. A strict lower limit is provided by the sum of all observed ions: $x_e \geq 2.2\ 10^{-9}$. The maximum value of $x(\text{M}^+)$, for which the ratio of the calculated to observed abundances lies between 0.5 and 2, is $4\ 10^{-9}$ in the case of Model 1 (Fig. 3a) and $2\ 10^{-8}$ in the case of Model 2. The predicted abundance of H$_3^+$ is below the upper limit derived by Black et al. (1990) and is in good agreement with the values estimated by van Dishoeck et al. (1992) from H$_2$D$^+$, in their high density case, especially when it is assumed that all nitrogen is in gas phase N$_2$. Model 1 predicts H$_2$D$^+$/H$_3^+$ and H$_2$DO$^+$/H$_3$O$^+$ ratios comparable to the observed ratios DCO$^+$/HCO$^+$ $\sim 5\ 10^{-3}$ and N$_2$D$^+$/N$_2$H$^+$ $\sim 2\ 10^{-3}$ (Table 9). In the case of Model 2, the calculated N$_2$D$^+$/N$_2$H$^+$ is higher due to the higher abundance of D, and the H$_2$D$^+$/H$_3^+$ is lower due to the lower abundance of HD. The model also predicts the abundance of NH$_3$, which is close to the value of $3\ 10^{-9}$ determined observationally by Schreyer et al. (1995), and that of NH$_2$D. It should be noted, however, that the current model does not include grain surface reactions, which may be significant for the production of NH$_3$ and its deuteration.

## 5.2 W 3 IRS5

The excitation analysis of §4 has shown that the density in W 3 IRS5 is slightly lower than that in NGC 2264, but the temperature higher. Moreover the observed $HCS^+/CS$ ratio of 3 $10^{-3}$ and the $N_2H^+$ abundance are lower, suggesting a higher electron fraction and/or a lower cosmic ray ionization rate. The adopted $CO/H_2$ ratio is also higher than toward NGC 2264. The model analysis for W 3 IRS5 cannot proceed in the same way as for NGC 2264, because no good determination of the $N_2H^+$ abundance is available. Although $N_2H^+$ 4–3 is most likely detected by van Dishoeck et al. (1992), the line falls at the edge of the spectrum and does not lead to a reliable column density. Moreover, the 5–4 line was not detected. On the other hand, $H_3O^+$ data are available for this source, in contrast with NGC 2264, which together with $HCS^+$ proved to be very useful in the analysis.

The abundance of $HCS^+$ can be written as:

$$\left[\frac{x(HCS^+)}{4.8\ 10^{-12}}\right] \sim 2.8 \frac{x(H_3O^+) + 1.2 x(HCO^+) + 11.9 x(H_3^+)}{x_e + 9.\ 10^{-5} x(O)}$$

Using the observational constraints on $HCS^+$, $H_3O^+$ and $HCO^+$, we can derive a relation between $x(H_3^+)$, $x(O)$ and $x_e$ from this expression.

The relation between $x_e$, $x(H_3^+)$ and $x(M^+)$ is given by:

$$x_e \sim x(H_3O^+) + x(HCO^+) + x(H_3^+) + x(M^+),$$

or

$$x_e \sim 2\ 10^{-9} + x(H_3^+) + x(M^+).$$

Together, these two formulae lead to an expression of $x(H_3^+)$ as a function of $x(O)$ and $x(M^+)$.

Further relations can be derived from

$$\left[\frac{x(H_3O^+)}{4.3\ 10^{-10}}\right] \sim \frac{2.6\ 10^6}{x_e}[x(H_2O)(x(HCO^+) + 1.7 x(H_3^+)) + 0.3 x(O) x(H_3^+)]$$

which leads to $x(H_2O)$ as a function of $x(O)$ and $x(M^+)$. In the same way an expression for $x(D)$ as a function of $x(O)$ and $x(M^+)$ is derived from:

$$\left[\frac{x(HCO^+)}{1.5\ 10^{-9}}\right] \sim \frac{x(N_2H^+) + 1.9x(H_3^+)}{\frac{x_e}{9.7\ 10^{-9}} + \frac{x(H_2O)}{2.5\ 10^{-6}} + \frac{x(D)}{6.3\ 10^{-6}}}$$

Assuming that $x(HD) = [D]/[H_2] - x(D)$, with $[D]/[H_2] = 2.8\ 10^{-5}$, another expression of $x(D)$ can be obtained from:

$$\left[\frac{x(DCO^+)}{3.7\ 10^{-12}}\right] \sim \frac{227.3}{\frac{x_e}{9.7\ 10^{-9}} + \frac{x(H_2O)}{2.5\ 10^{-6}}}[x(D)(x(HCO^+) + 7.7x(H_3^+)) + 10x(HD)x(H_3^+)]$$

The two expressions of $x(D)$ imply that:

$$\frac{x(O)}{3.9\ 10^{-5}} + \frac{x(M^+)}{8.2\ 10^{-9}} \sim 1.$$

The maximum value of $x(M^+)$ is then around $8.\ 10^{-9}$, corresponding to a low atomic oxygen abundance (Model 2). The values of these two parameters being fixed, we can obtain the abundances of $H_2O$, D and HD needed to reproduce the observations, as presented in Table 7: the abundance of $H_2O$ adopted in the modeling is in good agreement with the value derived by Phillips et al. (1992) from data of Cernicharo et al. (1990), and only 0.04 to 0.1% of deuterium is in atomic form. We also derive the abundance of $H_3^+$ from $x(O)$ and $x(M^+)$, and then $\zeta_H$ since:

$$x(H_3^+) \sim \frac{2.1\ 10^6 \zeta_H}{1 + \frac{x(O)}{5.7\ 10^{-4}} + \frac{x(C)}{2.3\ 10^{-4}} + \frac{x_e}{9.6\ 10^{-7}}}$$

In the case of Model 1 the values of $x(O)$ and $x(M^+)$ have been chosen in order to get a value of $\zeta_H$ close to the lower limit generally assumed in dark clouds.

As for NGC 2264, the sensitivity of the calculated ion abundances to the main parameters, $x(M^+)$, $x(H_2O)$ and $\zeta_H$ are shown in Fig. 3b, in the case of Model 2. The fraction of $N_2$ is chosen to give an abundance of $N_2H^+$ close to the observed upper limit: between 20 and 40% of the nitrogen is then in gas phase $N_2$, similar to the values suggested by van Dishoeck et al. (1992).

With the values of the parameters (Table 7), good agreement between the observed and calculated abundances is achieved for $HCO^+$, $DCO^+$, $N_2H^+$, $HCS^+$ and $H_3O^+$ (Table

Table 7: Model inputs for W 3 IRS5

| Parameter | Model 1 | Model 2 |
|---|---|---|
| $\zeta_{\rm H}$ (s$^{-1}$) | $10^{-17}$ | $2.5\ 10^{-17}$ |
| $x({\rm M}^+)$ | $3\ 10^{-9}$ | $8\ 10^{-9}$ |
| $x({\rm N}_2)$ | $4.4\ 10^{-5}$ | $2\ 10^{-5}$ |
| $x({\rm H}_2{\rm O})$ | $10^{-6}$ | $3\ 10^{-6}$ |
| $x({\rm O})$ | $2.5\ 10^{-5}$ | $\leq 2.\ 10^{-6}$ |
| $x({\rm C})$ | $\leq 2.\ 10^{-6}$ | $\leq 2.\ 10^{-6}$ |
| $x({\rm D})$ | $1.2\ 10^{-8}$ | $3\ 10^{-8}$ |
| $x({\rm HD})$ | $2.8\ 10^{-5}$ | $2.8\ 10^{-5}$ |

Table 8: Model fits for W 3 IRS5

| $x({\rm model})/x({\rm obs})$ | Model 1 | Model 2 |
|---|---|---|
| HCO$^+$ | 1.2 | 1.2 |
| DCO$^+$ | 1.0 | 1.0 |
| N$_2$H$^+$ * | 0.9 | 1.1 |
| N$_2$D$^+$ * | $3.6\ 10^{-4}$ | $6.2\ 10^{-4}$ |
| HCS$^+$ | 1.0 | 0.9 |
| H$_3$O$^+$ | 1.0 | 1.5 |
| H$_2$D$^+$ * | $1.2\ 10^{-4}$ | $3.1\ 10^{-4}$ |
| predictions | | |
| $x_e$ | $5.3\ 10^{-9}$ | $1.1\ 10^{-8}$ |
| $x({\rm H}_3^+)$ | $1.9\ 10^{-11}$ | $5.0\ 10^{-11}$ |
| $x({\rm H}_2{\rm DO}^+)$ | $7.8\ 10^{-13}$ | $1.2\ 10^{-12}$ |
| $x({\rm NH}_3)$ | $4.5\ 10^{-10}$ | $4.2\ 10^{-10}$ |
| $x({\rm NH}_2{\rm D})$ | $1.9\ 10^{-13}$ | $1.7\ 10^{-13}$ |

* Compared with the observed upper limit.

Table 9: Ratios of deuterated to hydrogenated ion abundances obtained in the models

| Ratio | NGC 2264 | | W 3 IRS5 | |
|---|---|---|---|---|
| | Mod. 1 | Mod. 2 | Mod. 1 | Mod. 2 |
| $N_2D^+/N_2H^+$ | $2.6\ 10^{-3}$ | $4.4\ 10^{-3}$ | $1.4\ 10^{-4}$ | $2.0\ 10^{-4}$ |
| $H_2D^+/H_3^+$ | $1.7\ 10^{-3}$ | $6.6\ 10^{-4}$ | $2.8\ 10^{-4}$ | $2.7\ 10^{-4}$ |
| $DCO^+/HCO^+$ | $3.9\ 10^{-3}$ | $3.2\ 10^{-3}$ | $2.1\ 10^{-3}$ | $2.1\ 10^{-3}$ |
| $H_2DO^+/H_3O^+$ | $2.9\ 10^{-3}$ | $1.2\ 10^{-3}$ | $1.8\ 10^{-3}$ | $1.9\ 10^{-3}$ |
| $NH_2D/NH_3$ | $3.7\ 10^{-4}$ | $1.6\ 10^{-4}$ | $4.2\ 10^{-4}$ | $4.1\ 10^{-4}$ |

8). The calculated values of $x(N_2D^+)$ and $x(H_2D^+)$ are much lower than the observed upper limits. The electron fraction is $x_e = (0.5 - 1.1)\ 10^{-8}$ and $x(M^+) = (3 - 8)\ 10^{-9}$. $M^+$ is the dominant ion, providing between 60 and 70% of the electrons. The predicted abundances of $H_3^+$, $NH_3$ and $NH_2D$ are one or two orders of magnitude lower than toward NGC 2264, due to the higher values of $x(CO)$ and $x(O)$ and the lower values of $\zeta_H$ and $x(N_2)$.

The predicted $H_2DO^+/H_3O^+$ ratio is close to the observed ratio $DCO^+/HCO^+ \sim 2.5\ 10^{-3}$, but is about one order of magnitude higher than the calculated $H_2D^+/H_3^+$, $N_2D^+/N_2H^+$ and $NH_2D/NH_3$ ratios (Table 9): these low values are mostly due to the low abundance of atomic deuterium adopted in the models required to reproduce the observed $DCO^+$ and $HCO^+$ abundances.

## 6   Conclusions and Discussion

We have observed submillimeter lines of several molecular ions and neutral species in two dense clouds, NGC 2264 and W 3 IRS5, which allow accurate determination of the physical conditions and abundances. Uncertainties remain due to the fact that possible density, temperature and abundance variations have been ignored, so that the derived $n(H_2)$, $T_{kin}$ and abundances are averaged over the beam and along the line of sight. The absolute abundances are furthermore uncertain due to lingering uncertainties in the $CO/H_2$ abundance ratio; the relative values should be more reliable, however.

In both clouds the lower limit on the electron fraction derived from the sum of the abundances of the observed ions is $x_e > 2\ 10^{-9}$ at densities of $(1 - 2)\ 10^6$ cm$^{-3}$. A more accurate determination of $x_e$ has been made, using a simple steady state chemical model in which the most important formation and destruction routes of the ions are taken into account, and in which as many species as possible are constrained by observations. Although there are still uncertainties in the values of the reactions rates, especially for dissociative recombination, we obtained agreement with the observations to within a factor of two. The possibility exists that a chemical steady state has not been reached in the observed sources, in which case comparison of observations with models are more difficult since the initial conditions and the age of the cloud are unknown. However, time-dependent models such as those of Millar et al. (1991) and Shalabiea & Greenberg (1995) show little variation in the major ion abundances and the electron fraction with time.

More elaborated models using a full blown gas phase network of more than 2000 reactions and including gas-grain interactions give comparable results: for a density $n(H_2) = 5\ 10^5$ cm$^{-3}$ and elemental gas-phase abundances close to those derived for NGC 2264 and W 3 IRS5, Shalabiea & Greenberg (1995) obtain $x_e \sim (1.5 - 3.5)\ 10^{-8}$ using a pure gas phase model or a dust/gas model (see their Model A). Similar results are presented by Le Bourlot et al. (1993) with a pure gas-phase model, as well as by Millar (1994, private communication).

We did not consider in this work the interaction between electrons and grains. As discussed e.g. by d'Hendecourt et al. (1985) this process leads in dense clouds to grains with an average charge of $-e$, due to the inefficient photoelectric effect. Because the grain abundance is typically $10^{-12}$, this results in a decrease in the gas phase electron fraction by only a few percent at most. Moreover, this situation may be modified if the internally produced ultraviolet photons due to the excitation of H$_2$ by cosmic rays are taken into account, as suggested by preliminary results of Cervetto et al. (1995): this ultraviolet radiation could be strong enough to give rise to a photoelectric effect sufficient to drive the charge distribution of grains toward positive values.

The ionization balance in dense clouds may also be modified by the presence of large molecules such as PAHs. In particular, the density of electrons may be reduced and negative PAH$^-$ ions may contain most of the negative charge if the total PAH abundance and the thermal sticking coefficient of electrons on neutral PAH molecules are high (Omont 1986, Lepp & Dalgarno 1988, Allamandola et al. 1989). This means that the electron fractions derived in the preceding section are upper limits, but that the degree of ionization could be higher, typically of the order of the total fraction of PAH which is generally assumed to be $x_{PAH} \sim 10^{-7} - 10^{-6}$, for PAH molecules containing between 1 and 10% of the total carbon and formed with about 80 C atoms. The reduced electron fraction would imply an increase in the abundances of those ions which are predominantly destroyed by dissociative recombination – H$_3$O$^+$, HCS$^+$, HCO$^+$ and $DCO^+$ in our case – , although the (slower) destruction by mutual neutralization with $PAH^-$ ions diminishes the effect. Thus, the numerical results of Section 5 could be slightly modified if PAH molecules are included in the modeling. However, the large uncertainties in the total PAH abundance in dense clouds and in the rates involving these species prevent clear conclusions. In particular, Mendoza-Gómez et al. (1995) argue that PAH molecules do not survive accretion onto grains inside dense clouds.

In summary for dense clouds with $n(H_2) \sim 10^6$ cm$^{-3}$, the inferred electron fraction lies in the range $2\ 10^{-9} < x_e < 4\ 10^{-8}$. Ionized metals such as $Mg^+$ or $Fe^+$ could provide up to 60% of the electrons. These values of $x_e$ are consistent with those commonly used in the literature and represented by the simple formula $x_e \sim 10^{-5}\ n(H_2)^{-1/2}$ (McKee 1989). Comparison of our results to those obtained in lower density, colder clouds (Schilke et al. 1991), does not show a clear dependence of $x_e$ on density between $10^4$ and $10^6$ cm$^{-3}$, however. In fact, the analysis of the preceeding section shows that other parameters which can fluctuate from cloud to cloud also play a rôle.

# 7 Acknowledgments

The authors are grateful to Remo Tilanus and Fred Baas for their support during the JCMT observing runs, and to Katharina Schreyer for sharing her JCMT data on NGC 2264 with us. They thank Tom Millar and Osama Shalabiea for many helpful discussions concerning chemical models. One of us (CB) would like to acknowledge E.F. van Dishoeck and her group for their hospitality in Leiden. This work was sponsored by a PIONIER grant from the Netherlands Organization for Scientific Research (NWO).